\documentclass[useAMS,usenatbib]{mn2e}

\usepackage{epsfig}

\newcommand{\nodata}{$\dots$}
\newcommand{\etal}{\mbox{et al.}}
\newcommand{\ergcms}{erg cm$^{-2}$ s$^{-1}$}
\newcommand{\ergs}{erg s$^{-1}$}

\newcommand{\msun}{M$_{\odot}$}

\newcommand{\apj}{{\it ApJ}}

\newcommand{\aap}{{\it A\&A}}

\newcommand{\mnras}{{\it MNRAS}}

\newcommand{\nat}{{\it Nature}}

\newcommand{\chandra}{{\it Chandra}}

\newcommand{\xmm}{{\it XMM-Newton}}

\newcommand{\swift}{{\it Swift}}

\newcommand{\wdpulsar}{\mbox{CXOU J164710.2-455216}}

\newcommand{\gogus}{G\"{o}\u{g}\"{u}\c{s}}

\begin{document}

\title[Exciting a Magnetar's Fields]{Exciting the Magnetosphere of the Magnetar \wdpulsar\ in Westerlund 1}

\author[M. P. Muno \etal]{
M. P. Muno,$^{1}$ B. M. Gaensler,$^{2,3}$ J. S. Clark,$^{3,4}$ 
R. de Grijs,$^{5}$
D. Pooley,$^{6,7}$ \newauthor
I. R. Stevens,$^{8}$ \& S. F. Portegies Zwart$^{9,10}$\\
$^{1}${Space Radiation Laboratory, California Institute of Technology, Pasadena, CA 91125; mmuno@srl.caltech.edu}\\
$^{2}${School of Physics A29, The University of Sydney, NSW 2006, Australia}\\
$^{3}${Harvard-Smithsonian Center for Astrophysics, 60 Garden St. Cambridge, MA 02138}\\
$^{4}${Department of Physics \& Astronomy, The Open University, 
Walton Hall, Milton Keynes, MK7 6AA, UK}\\
$^{5}${Department of Physics \& Astronomy, The University of 
  Sheffield, Hicks Building, Hounsfield Road, Sheffield S3 7RH, U.K.}\\
$^{6}${Chandra Fellow}\\
$^{7}${Astronomy Department, University of California at Berkeley, 
 601 Campbell Hall, Berkeley, CA 94720, USA}\\
$^{8}${School of Physics \& Astronomy, University of Birmingham, Edgbaston, Birmingham B15 2TT, UK}\\
$^{9}${Astronomical Institute 'Anton Pannekoek'
        Kruislaan 403, 1098SJ Amsterdam, the Netherlands}\\
$^{10}${Section Computational Science 
        Kruislaan 403, 1098SJ Amsterdam, the Netherlands}
}

\date{Accepted 2007 March 15. Received 2007 February 13; in original form 2006 November 25}

\pagerange{\pageref{firstpage}--\pageref{lastpage}} \pubyear{2007}

\maketitle

\label{firstpage}

\begin{abstract}
We describe \xmm\ observations taken 4.3 days prior to and 1.5 days
subsequent to two remarkable events that were detected with 
\swift\ on 2006 September 21 from the candidate magnetar
\wdpulsar: (1) a 20 ms burst with an energy of $10^{37}$ erg (15--150 keV), 
and (2) a rapid spin-down (glitch) with $\Delta P/P \sim -10^{-4}$. We
find that the luminosity of the pulsar increased by a factor of 100 in
the interval between observations, from $1\times10^{33}$ to
$1\times10^{35}$ \ergs\ (0.5--8.0~keV), and that its spectrum
hardened.  The pulsed count rate increased by a factor of 10
(0.5--8.0~keV), but the fractional rms amplitude of the pulses
decreased from 65 to 11 per cent, and their profile
changed from being single-peaked to exhibiting three
peaks. Similar changes have been observed from other magnetars in
response to outbursts, such as that of 1E 2259+586 in 2002 June.  
We suggest that a plastic deformation of the
neutron star's crust induced a very slight twist in the external
magnetic field, which in turn generated currents in the magnetosphere
that were the direct cause of the X-ray outburst.
\end{abstract}


\section{Introduction}

Young, isolated neutron stars come in a variety of manifestations,
including ordinary radio pulsars, compact
central objects in supernova remnants, soft gamma repeaters (SGRs),
and anomalous X-ray pulsars (AXPs). The latter two classes of source
share long rotational periods ($P$=5--10 s), rapid spin-down rates
($\dot{P}$$\ga$$10^{-12}$ s s$^{-1}$), X-ray luminosities 
($L_{\rm X}$$\ga$$10^{33}$ \ergs) that exceed their spin-down power, and
the frequent production of second-long soft gamma-ray bursts \citep{wt06}. 
These properties suggest that they are magnetars, 
neutron stars powered by the unwinding of extremely strong 
($B$$\ga$$10^{15}$ G) internal magnetic fields \citep{td95,td96}.

The phenomenology associated with magnetars is thought to be driven by
how the unwinding internal fields interact with the crusts of the
neutron stars, which in turn determines the geometries of the external
magnetic fields (Thompson \& Duncan 1995, 1996; Thompson, Lyutikov, \& 
Kulkarni 2002)\nocite{td95,td96,tlk02}. In some cases, the crusts
respond to the unwinding fields plastically, and the energy is
gradually deposited into the magnetospheres.  This causes transient
`active periods,' in which the persistent fluxes increase on
timescales of weeks to years \citep{woo04,got04}.  Fractures may also
occur in the crust, which generate waves in the external fields, and
in turn produce sudden soft gamma-ray `bursts' with energies up to
$10^{41}$ erg (\gogus\ \etal\ 2001; Gavriil, Kaspi, \& Woods
2002)\nocite{gog01,gkw02}. In the most extreme cases, instabilities
can rearrange the entire external magnetic field, producing `giant
flares' with energies of $10^{44} - 10^{46}$ erg
\citep{hur99,pal05,hur05}.  Finally, changes in the coupling between 
the bulk of the crust and a superfluid component appear to change
the crust's angular momentum, as is suggested by both secular variations in 
the spin down rates on time-scales of weeks
\citep{gk04,woo06} or sudden, day-long episodes of spin-up (`glitches') 
or spin-down \citep{woo99,gk03,dal03,kas03,woo04}.
Unfortunately, the frequent, sensitive
monitoring observations that are required to identify transient
active periods, to detect bursts, and to track the rotation of these
pulsars have not always been available. Therefore, in many cases the 
causal connections between these phenomena have been unclear 
\citep[e.g.,][]{gk03,woo05}. 

Here we report \xmm\ observations of the 10.6 s X-ray pulsar,
\wdpulsar\ \citep{mun06a}, that bracketed a series of 
events that occurred near 2006 September 21. Near this time, 
\swift\ detected
a soft gamma-ray burst \citep{kri06} and a glitch with 
$\Delta P/ P \sim -10^{-4}$ \citep{isr07}. 
These events confirm our original hypothesis that this source is a magnetar
\citep{mun06a}. We find that during the interval
between our two \xmm\ observations, there were also dramatic changes in the
luminosity, spectrum, and pulse profile of \wdpulsar. We compare these
to changes observed during active periods from other magnetars, and 
discuss the implications for the interaction between the magnetic fields
and crusts of the these neutron stars.

\begin{figure}
\centerline{\psfig{file=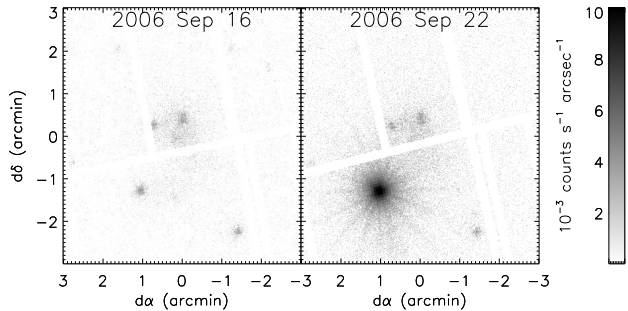,width=\linewidth}}
\caption{Images of the counts received by the EPIC-pn in the 0.5--10 keV band 
on 2006 September 16 (left) and 22 (right). The images are centred on the
core of the star cluster Westerlund 1 
($\alpha$, $\delta$ = 251$.\!^{\rm h}$76792 --45$.\!^\circ$84972 [J2000]). 
In addition to the AXP \wdpulsar, also visible in the images are three bright 
OB/WR stars. The blank strips in the image are gaps between the chips 
in the detector array.
}
\label{fig:img}
\end{figure}

\section{Observations}

As part of the guest observer programme, \xmm\ observed \wdpulsar\ for 
46 ks starting on 2005 September 16 at 18:59:38 (UTC). Fortuitously,
4.3 days later, on 2006 September 21 at 01:34:53 (UTC), the \swift\ Burst
Alert Telescope (BAT) detected a 20 ms burst from the direction of 
Westerlund 1 \citep{kri06}, with an energy of $3\times10^{37}$ erg 
(15--150 keV; for a distance $D$=5 kpc; Clark \etal\ 2005). 
In response, the director
of \xmm\ carried out an observation lasting 30 ks beginning 
1.5 days later on 2006 September 22 at 12:40:27 (UTC). We analysed the \xmm\
observations in order to study changes in the 
X-ray flux, spectrum, and pulse profile.

We analysed the data taken with the European Photon Imaging Camera 
(EPIC).
For most of the timing and spectral analysis, we 
used data taken with 73.4 ms time resolution using the pn array. The 
data from the MOS arrays were taken with 2.4 s time resolution, which was
inadequate for studying the profile of this 10.6 s pulsar. Moreover,
the data suffered from pile-up during the second observation, when 
the source was bright (see below). Therefore, we only used the MOS data 
to generate spectra for the first observation. 

We processed the observation data files
using the standard tools ({\tt epchain} and {\tt emchain}) 
from the Science Analysis Software version 7.0. The events
were filtered in the standard manner, 
and we adjusted the arrival times of the events to the Solar System 
barycentre.
Images from the EPIC-pn data are displayed in Figure~\ref{fig:img}.
Comparing the data from before and after the \swift\ 
burst, we find that \wdpulsar\ increased in count rate by a factor of 
80 (0.5--8.0 keV).

\begin{figure}
\centerline{\psfig{file=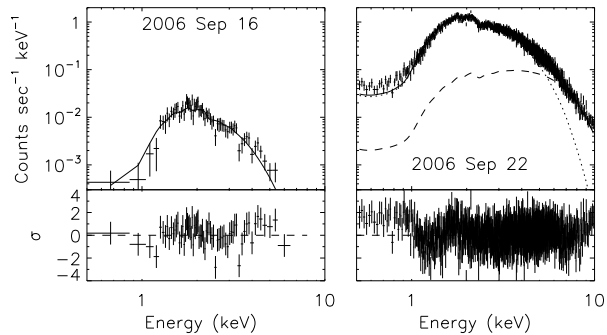,width=\linewidth}}
\caption{Phase-averaged spectra of \wdpulsar\ taken on 2006 September 16 
and 22 (top panels), in units of detector counts. Models for the spectra 
are shown
with a solid line: a single absorbed blackbody on September 16,
and two absorbed blackbodies on September 22. For the latter
spectrum, the cool and hot blackbodies are indicated with the dotted and 
dashed lines, respectively. The bottom panels 
display the difference between the data and the models, in units of the
1$\sigma$ uncertainty on the data. There 
are systematic residuals at low energies in the September 22 spectrum, but 
these are not significant enough to affect the overall model.
}
\label{fig:spec}
\end{figure}

Next, we extracted pulse-phase-averaged spectra from within 15\arcsec\ of the 
location of \wdpulsar\ 
($\alpha$, $\delta$ = 251$.\!^{\rm h}$79250, --45$.\!^\circ$87136 [J2000]).
Estimates of the background were extracted from a 30\arcsec\ circular region
that was located 1\farcm5 west of the source region. 
We obtained the detector response and effective area using standard tools 
({\tt rmfgen} and {\tt arfgen}). The EPIC-pn spectra are displayed in 
Figure~\ref{fig:spec}.

We modeled these spectra using {\tt XSPEC} version 12.2.1. 
We first assumed that the spectra could be described as blackbody emission
absorbed by interstellar gas and scattered by dust. This model was 
acceptable for the 
observations before the burst on September 16 ($\chi^2/\nu$ = 59.4/67),
but was inconsistent with 
the data from September 22 ($\chi^2/\nu$ = 2255/1136). For the later
observation, we could model the spectrum with two continuum components,
either the sum of two blackbodies, or a blackbody plus power law continuum.
We assumed that the interstellar absorption column toward the source
did not change between observations. The spectral parameters, fluxes, 
and luminosities for the above models are listed in Table~\ref{tab:bb}. 
For completeness,
we also list parameters from models of the spectra taken with \chandra\
during 2005 May and June \citep{mun06a}.

\begin{table}
\caption{Spectral Models for \wdpulsar\label{tab:bb}}
\begin{tabular}{@{}lccc@{}}
\hline
 & 2005 & \multicolumn{2}{c}{2006} \\
 & May--Jun & Sep 16 & Sep 22 \\
\hline
\multicolumn{4}{c}{Two Blackbodies}\\
$N_{\rm H}$ ($10^{22}$ cm$^{-2}$) & 1.28 & 1.28 & 1.28(2) \\
$kT_1$ (keV) & $0.60(1)$ & $0.54(1)$ & $0.67(1)$ \\
$A_{\rm bb,1}$ (km$^2$) & $0.09(1)$ & $0.08(1)$ & $3.62(2)$ \\ 
$kT_2$ (keV) & \nodata & \nodata & $1.7(1)$ \\
$A_{\rm bb,2}$ (km$^2$) & \nodata & \nodata & $0.021(6)$ \\
$F_{\rm X}$ ($10^{-13}$ \ergcms) & 2.3 & 1.5 & 215.7 \\
$L_{\rm X}$ ($10^{33}$ \ergs) & 1.4 & 1.0 & 109.7\\
\hline
\multicolumn{4}{c}{Blackbody Plus Power Law}\\
$N_{\rm H}$ ($10^{22}$ cm$^{-2}$) & 1.44 & 1.44 & 1.44(1) \\
$kT_1$ (keV) & $0.58(2)$ & $0.52(1)$ & $0.68(1)$ \\
$A_{\rm bb,1}$ (km$^2$) & $0.11(1)$ & $0.11(1)$ & $2.87(3)$ \\ 
$\Gamma$ & \nodata & \nodata & $2.07(4)$ \\
$N_{\Gamma}$ ($10^{-3}$ cm$^{-2}$ s$^{-1}$ keV$^{-1}$) & \nodata & \nodata & $3.7(9)$ \\
$F_{\rm X}$ ($10^{-13}$ \ergcms) & 2.3 & 1.5 & 214.1 \\
$L_{\rm X}$ ($10^{33}$ \ergs) & 1.5 & 1.1 & 130.3\\
\hline
\end{tabular}
The reduced chi-squared for both joint fits were 1423/1298. 
The interstellar absorption was assumed not to have 
changed over the course of these observations.
To compute the area of the blackbody emission, we assumed $D$=5 kpc.
$N_{\Gamma}$ is the photon flux density of the power law at 1 keV.
Uncertainties are 1$\sigma$, for one degree of freedom. Fluxes
are in the 0.5--8.0 keV band.
\end{table}

For both models, we found that the luminosity
was a factor of 100 higher (0.5--8.0 keV) 1.5 days after the burst 
than it was 4.3 days before the burst. The increase in flux was 
largely because the area of 
the $\approx$0.5 keV blackbody increased from 0.1 km$^2$ before the burst 
to $\approx$3 km$^2$ after the burst. It also resulted from the
prominence of the hard component after the burst. Modeled as a 
$kT$=1.7 keV blackbody, it produced 26 per cent of the observed flux 
on 2006 September 22 (18 per cent of the absorption-correction flux). Modeled
as a $\Gamma$=2.07 power law, it produced 50 per cent of the observed flux 
(70 per cent of the intrinsic flux; 0.5-8.0 keV). If we add these components 
to our models for the
spectra taken on 2006 September 16, we find that their fractional 
contribution to the observed flux was lower: $<$15 per cent for the blackbody, 
and $<$35 per cent for the power law. 

To identify pulsations from \wdpulsar, we computed 
Fourier periodograms using the Rayleigh statistic. 
(A search for pulsations from other
point sources in the field revealed no other pulsars.)
This provided an initial estimate of the pulse period, which we then 
refined by computing pulse profiles from non-overlapping 5000 s intervals 
during each observation, measuring their phases by cross-correlating them
with the average pulse profile from each observation, and modeling the 
differences between the assumed and measured phases using first-order 
polynomials. The best-fitting periods were 10.61065(7) s and 
10.61064(8) s for 2006 September 16 and 22, respectively. These values
are within 1.5$\sigma$ of the periods measured in 2005 May and June, 
10.6112(4)~s and 10.6107(1)~s, respectively \citep{mun06a}.
The reference epochs of the pulse maxima for the two observations in 
2006 September were 53994.786313(2) and 54000.526588(1) (MJD, Barycentre
Dynamical Time). Monitoring observations taken with \swift\ reveal that
a glitch with a fractional period change of $\Delta P/P \sim -10^{-4}$ 
occurred between these two observations; a discussion
of this result is presented in \citep{isr07}.

\begin{figure}
\centerline{\psfig{file=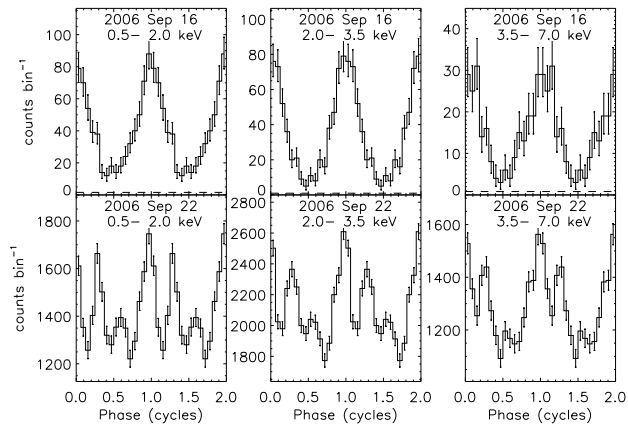,width=\linewidth}}
\caption{Pulse profiles of \wdpulsar\ taken on 2006 September 16 (top
panels) and 2006 September 22 (bottom panels), and in three energy 
bands: 0.5--2.0 keV
(left panels), 2.0--3.5 keV (middle panels), and 3.5--7.0 keV (right panels).
Two identical cycles are repeated in each panel. The dashed line in the top
panel represents the background count rate.}
\label{fig:prof}
\end{figure}

We used these ephemerides to compute the pulse profiles in the full band 
of 0.5--8.0 keV, and three sub-bands: 0.5--2.0 keV, 2.0--3.5 keV, and 
3.5--7.0 keV. 
The root-mean-squared (rms) amplitudes of the pulsations in the 
full band (0.5--8.0 keV) increased from $0.02$ count s$^{-1}$ before the 
burst, to $0.29$ count s$^{-1}$ after 
the burst. At the same time, the fractional rms amplitudes declined from 64
per cent 
before the burst to 11 per cent after the burst. Moreover, the pulse profile
changed dramatically after the burst, as can be seen in the profiles from
the sub-bands displayed in Figure~\ref{fig:prof}. 
Before the burst, the
pulse at all energies was single peaked, and the differences in the
pulse profile as a function of energy are not very pronounced. After
the burst, the pulse in the full band displayed three distinct peaks,
and a dependence on energy developed. Specifically, in the 3.5--7.0 keV band,
the third peak was absent and the flux between the first two peaks (phases
0.1--0.3) was larger, so that the overall profile
was more sinusoidal at high energies than at low.

We examined whether phase-resolved spectroscopy could provide any
insight into the origin of the pulses. Unfortunately, \wdpulsar\
was too faint on 2006 September 16 to generate spectra for all but
the peak of the pulse. We did examine phase-resolved spectra for 
2006 September 22, but found no systematic trend relating
the spectral parameters with the intensity as a function of phase.


Finally, we searched for bursts by examining the time series of events 
recorded by the EPIC-pn. We found no evidence for bursts producing more 
than 4 counts within the 73.4 ms frame time, which placed an
upper limit to their observed fluence of $3\times10^{-11}$ erg cm$^{-2}$ 
\citep[for a $\Gamma$=1.8 power law;][]{kri06}, or 
an energy of $<$$2\times10^{35}$ erg (0.5--8.0 keV; $D$=5 kpc).

\section{Discussion}

In the 5.8 days between our two \xmm\ observations of \wdpulsar, a 
remarkable set of events occurred. 
First, the phase-averaged
luminosity of \wdpulsar\ increased by a factor of $\sim$100, from
$L_{\rm X} = 1\times10^{33}$ to $L_{\rm X} = 1\times10^{35}$ \ergs\ 
\citep[0.5--8.0 keV; Fig.~\ref{fig:img};][]{ci06}, and the
spectrum hardened (Table~\ref{tab:bb}). Energetically, this is the 
most important feature of this active period. In the 1.5 days after 
the burst, if we conservatively assume the persistent flux from 
\wdpulsar\ was constant, the total energy released was $\sim$$10^{40}$ erg
(0.5--8.0 keV).
Second, a 20 ms long burst with an energy of $3\times10^{37}$ erg (15--150
keV) was detected from this source with the BAT on board \swift\
\citep{kri06}.
Third, a glitch 
was observed in the spin period of the pulsar, with 
$\Delta P/P \sim -10^{-4}$ \citep{isr07}. 
Fourth, the pulse 
profile changed from having a simple, single-peaked structure, to 
exhibiting three distinct peaks with pronounced energy dependence 
(Fig.~\ref{fig:prof}). Similar changes in the fluxes, spectra, and 
timing properties of magnetars have been observed before, but the 
combination observed from \wdpulsar\ is unique.

It is common for the persistent luminosities of magnetars to vary on time
scales of weeks to years. The persistent luminosities from the
SGRs 1900+14 \citep{woo01} and 1806--20 \citep{woo06} and the bright AXPs 
1E 1048.1--5937 \citep{gk04,tie05} and 1E 2259+586 \citep{woo04} have been
observed to vary by factors of 2--3 around $\sim$$10^{34}-10^{35}$ \ergs\
(0.5--10 keV). 
The luminosities of SGR 1627--41 \citep{kou03} and the transient AXP 
XTE J1810--597 \citep{ibr04,got04} have been observed to increase by 
factors of 100, from $\sim$$10^{33}$ to $\sim$$10^{35}$ \ergs\ (0.5--10 keV). 
The larger luminosities,
$\sim$$10^{35}$ \ergs, appear to be a rough upper envelope for the 
persistent 0.5--8.0 keV fluxes of magnetars 
(not counting bursts and giant flares). 
Indeed, the active period from \wdpulsar\ also had $L_{\rm X} \approx 10^{35}$ 
\ergs\ (0.5--8.0 keV). This persistent flux is generally assumed to be 
produced because the unwinding internal fields induce  
gradual, plastic deformations in the crust and external magnetic fields, 
which in turn heats the surface or magnetosphere
\citep{td95, td96}. Therefore, the increase in the flux
from \wdpulsar\ demonstrates that either the unwinding 
of the internal fields, or the response of the crust to that unwinding, is
intermittent and can activate in $\la$5 days. 

The active periods from magnetars are often accompanied by
second-long bursts. These bursts are the hallmarks of SGRs, and during
their active periods hundreds will occur over the course of a year with
energies of up to $10^{41}$ erg
\citep[2--60 keV;][]{gog01}.
The bursts detected from AXPs have all been weaker, with peak energies
of $\la$$10^{38}$ erg (2--60 keV). In the AXPs XTE J1810--597
\citep{woo05} and 1E 1048.1--5937 
(Gavriil, Kaspi, \& Woods 2006)\nocite{gkw06}, the bursts that have
been detected are infrequent and relatively isolated.  In 1E 2259+586
\citep{woo04}, a series of bursts were detected during an 11 ks
observation that occurred within 7 days of the start of an active
period in 2002 June. The burst detected from \wdpulsar\ resembles
those from 1E 2259+586, in that it occurred very near the start of an
active period. The energy of the burst ($3\times10^{37}$ erg; 15--150 keV) 
is trivial compared to that released as persistent flux ($\ga$$10^{40}$ erg;
0.5--8.0 keV), so it is probably not a trigger, but a symptom of the 
active period.
Under the magnetar model, the bursts that accompany the 
active periods are caused by fractures that occur in the crust.  These
fractures inject into the magnetosphere currents that are unstable to 
to wave motion, which quickly generates hot, 
X-ray emitting plasma \citep{td95,td96}.  It is reasonable to expect 
that such fractures
would be stronger and occur more frequently when the persistent flux
is higher, because the crust is already under stress.

Variations in the spin-down rates have been 
observed from several luminous ($L_{\rm X} \ga 10^{34}$ \ergs; 0.5--8.0 keV) 
magnetars. Torque variations have been detected from 
1E 1048.1--5937 \citep{gk04} and SGR 1806--20 \citep{woo06}, in 
association with their active periods. Sudden period changes 
have been seen in three cases.
Two glitches have been detected from 1RXS J170849--400910
with $\Delta P/P \sim -1\times10^{-6}$ and $-6\times10^{-6}$ 
\citep{gk03,dal03}. Neither were associated with active periods, but the 
monitoring observations were sparse, so one could have
been missed \citep{dal03}. 
One glitch accompanied the 2002 June active period of 1E 2259+586
in which the spin period decreased by $\Delta P/P \sim -10^{-6}$ 
\citep{kas03,woo04}. Finally, a dramatic episode of spin-down occurred near 
the time of a $10^{44}$ erg (3--100 keV) giant flare from SGR 1900+14,
with $\Delta P/P \sim 10^{-4}$ \citep{woo99}.
This is of comparable magnitude to the glitch from \wdpulsar, albeit
of the opposite sign \citep{isr07}.

The glitch appears to have been a major energetic component of the outburst
from \wdpulsar. Glitches are ascribed to sudden changes in the 
moments of inertia of the neutron stars that occur when crustal 
movements change how superfluid in the interior is coupled to the 
bulk of the crust \citep[e.g.,][]{dal03,kas03}. The change in rotational 
energy during the glitch, assuming most of the star rotates as a solid 
body, is on order $\Delta E_{\rm rot} \sim I \Omega \Delta\Omega$, 
where $I$$\sim$$10^{45}$ g cm$^2$ is the moment of inertia of a 
neutron star with mass $M$$=$$1.4$ \msun\ and radius $R$$=$$1$ km. For  
\wdpulsar\ $\Omega$$=$$0.6$ rad s$^{-1}$ and $\Delta\Omega$$=$$6\times10^{-5}$
rad s$^{-1}$, so $\Delta E_{\rm rot}$$\sim$$10^{40}$ erg. However, a 
larger input of energy into the stellar interior may be required to unpin 
the superfluid vortices and initiate the glitch, $\sim$$10^{42}$ erg
\citep[e.g.,][]{le96,tho00}. In contrast, the radiative output of \wdpulsar\
in the first week of this active period was only $\sim$$10^{40}$ erg 
(0.5--8.0 keV). Whereas for the giant flare from SGR 1900+14 and the 2002
June active period from 1E 2259+586 it appeared 
that most of the energy was radiated away from 
the magnetosphere \citep{tho00, woo04}, for \wdpulsar\ most 
of the energy was probably input into the interior of the neutron star.

The change in the pulse profile of \wdpulsar\ is also difficult to understand
from an energetic standpoint. Changes in the
qualitative shape of the pulse profiles (as opposed to changes in 
the pulsed fraction) have only been seen previously from three sources.
For 1E 2259+586, the profile before the 2002 June burst exhibited 
two distinct peaks, whereas after the burst the phases between the 
peaks contained more flux, so that part of the profile resembled a single 
plateau of emission \citep{woo04}. This change is minor compared
to that from \wdpulsar\ in Figure~\ref{fig:prof}. Large changes in the 
harmonic structure of the pulse profile have only been observed 
in response to the giant flares from SGRs. 
For SGR 1900+14 the profile had three peaks before 
the flare in 1998, and a single peak during and after \citep{woo01}. 
For SGR 1806--20, the opposite change occurred in 2004: it shifted from 
having a simple, single-pulsed profile to having multiple peaks \citep{woo06}.

For the SGRs, the changes in the pulse profiles are thought to 
occur because the multipole structure of the external magnetic
fields are rearranged. This is reasonable, because the giant flares 
release a significant fraction of the energy in the 
external fields. For a dipole, this would be 
$E_B \approx \frac{1}{12} B_{\rm ext}^2 R^3 \sim 10^{45}$ G, where we take
$B_{\rm ext}$$\sim$$10^{14}$ G, and $R$$\sim$10 km \citep{woo99,hur05}. 
However, for \wdpulsar, and to a lesser degree for 1E 2259+586, it is
unreasonable to suggest that active periods releasing only $\sim$$10^{40}$ erg 
of X-rays resulted from a significant rearrangement of the exterior 
magnetic fields.

Instead, we suggest that a change occurred in the distribution of
currents in the magnetosphere.  We hypothesize that the emission in
quiescence is thermal emission from the cooling neutron star, which
emerges through a hot spot where the opacity of the highly-magnetized
atmosphere is lowest \citep{hh98}. A single hot spot on the
surface could explain the single-peaked, fully modulated
($\approx$70 per cent rms) pulse in quiescence (\"{O}zel, Psaltiz, \& Kaspi
2001)\nocite{opk01}. We suggest that the active period was initiated
when a very small twist was imparted to the magnetic field by plastic
motions of the crust. Currents formed to compensate for this twist,
which heated the surface of the star and resonantly scattered the
emission from its surface (Table~\ref{tab:bb}).
Both of these would contribute to
creating the complex pulse profile \citep{tlk02}. 
If our scenario is correct, when this source returns to quiescence, the 
pulse should regain its single-peaked profile.

\section{Conclusions}

We have examined the X-ray luminosity, spectrum, and pulse profile of 
\wdpulsar\ before and after an interval during which \swift\ detected 
a soft gamma-ray burst and a timing glitch from the source. 
The energy radiated from the exterior was too small to have 
resulted from a significant rearrangement of the external magnetic fields 
of \wdpulsar. Instead, the dramatic change in the pulse profile indicates 
that the underlying emission mechanism changed. Before the burst, the 
X-ray emission was probably powered by the thermal energy of the star, 
whereas afterwards it was powered by currents in the magnetosphere.
Moreover, the glitch required an energy at least as large as the 
energy released as X-rays, $\ga$$10^{40}$ erg, which suggests that much 
of the energy of this event was input into the interior of the neutron star. 
Future X-ray observations of this source will reveal the duration and duty 
cycle of this active period, which would constrain the amount of energy
input into the interior. This could help answer why the 
emission, which is thought to be produced as the internal fields of magnetars 
unwind, can remain inactive for years and then suddenly turn on in a few days.

\section*{Acknowledgments}
We thank N. Schartel for providing the discretionary observation, 
G. Israel for sharing 
the results of the \swift\ observations, and the referee for helpful
comments.  MPM was supported by the NASA {\it XMM} 
Guest Observer Facility; BMG
by a Federation Fellowship from the Australian Research Council 
and an Alfred P. Sloan Research Fellowship; and SFPZ by 
the Royal Dutch Academy of Arts and Sciences.

\label{lastpage}

\end{document}